\input epsf
\documentstyle[12pt,aasms4]{article}
\doublespace
\def\wig#1{\mathrel{\hbox{\hbox to 0pt{%
          \lower.5ex\hbox{$\sim$}\hss}\raise.4ex\hbox{$#1$}}}}

\newcommand{\beq}{\begin{equation}}
\newcommand{\eeq}{\end{equation}}
\newcommand{\eeql}[1]{\label{eq:#1}\end{equation}}

\newcounter{compteur}

\def\bib{\par\noindent\hangindent=3mm\hangafter=1}

\def\be{\begin{equation}}
\def\ee{\end{equation}}
\def\m2i{M_{2,\rm i}}

\def\msol{M_\odot}
\def\lsun{L_{\odot}}

\def\te{T_{\rm eff}}

\def\simgr{\,\hbox{\hbox{$ > $}\kern -0.8em \lower 1.0ex\hbox{$\sim$}}\,}
\def\simle{\,\hbox{\hbox{$ < $}\kern -0.8em \lower 1.0ex\hbox{$\sim$}}\,}

\begin{document}

\title{\bf Deuterium-burning in substellar objects}

\author{ {\sc G. Chabrier$^{1}$, I. Baraffe$^{1}$,
F. Allard$^{1}$ \and P. Hauschildt$^{2}$ }}

$^{1}$ {Ecole Normale Sup\'erieure de Lyon, C.R.A.L (UMR 5574 CNRS),\\
\indent 69364 Lyon
 Cedex 07, France\\ \indent  email: chabrier, ibaraffe, fallard @ens-lyon.fr

$^{2}$Dept. of Physics and Astronomy and Center for Simulational Physics,
\\  \indent University of Georgia Athens, GA 30602-2451 \\  \indent email:
yeti@hobbes.physast.uga.edu}

\begin{center}
$\rm\underline{Submitted\ to}$: {\sl ApJ Letters}

\bigskip
\end{center}
\bigskip
\bigskip






\begin{abstract}
 
We consider the depletion of primordial deuterium in the interior of substellar objects as a function of mass, age and absolute magnitude in several photometric passbands. We characterize potential spectroscopic signatures of deuterium in the lines of deuterated water HDO. These results will serve as a
useful, independent diagnostic to characterize the mass and/or the age of young substellar objects, and to provide an independent age determination of very young clusters. These results can serve to identify objects at the deuterium-burning limit and to confront the theoretical prediction that D-burning is a necessary condition to form star-like objects.

\bigskip
\bigskip

Subject headings : stars: low mass, brown dwarfs --- stars: interiors ---  stars: atmosphere ---  planetary systems ---  stars: evolution
\end{abstract}


\section{Introduction}
The search for very low mass stars and substellar objects (SSOs) progresses nowadays at an astonishing pace. Many brown dwarfs (BDs) have now been discovered either in the field or in young clusters, confirming their predicted general evolution in various observable diagrams (see e.g. Basri 2000; Chabrier \& Baraffe 2000 for recent reviews). This evolution relies on both the properties of dense degenerate matter in the interior, which determines the mechanical properties (mass, radius, central density and temperature), and on the complex thermochemistry and radiation processes taking place in the atmosphere, which characterize the thermal properties (effective temperature, spectral distribution) and thus the observable signature, magnitude and color.
In the meantime several tens of exoplanets orbiting G-M stars have now been discovered by Doppler techniques, spaning a range from about 11 Jupiter mass ($\simeq 0.011\,\msol$) down to about a Saturn mass ($\simeq 3\times 10^{-4}\,\msol$).
By definition SSOs never reach thermal equilibrium since their core temperature is not hot enough to sustain hydrogen-burning and prevent further gravitational contraction. For this reason they do not define a main sequence, as stars do, and age adds another degree of freedom besides mass for a given observational signature.

To circumvent this difficulty, the abundance of lithium in the atmosphere of low-mass objects has been suggested as an independent diagnostic to determine the mass and/or age of an object below the bottom of the main sequence,  the so-called
"lithium-test" (Rebolo et al. 1992; Basri et al. 1996). The temperature for lithium-burning through the $^7Li(p,\alpha)^4He$ reaction is lower than the one required for proton fusion. Objects below 0.06 $\msol$ (Chabrier \& Baraffe 1997) are not massive enough - and thus not hot enough - to host lithium burning and thus retain their initial lithium abundances. Furthermore, these objects are fully convective and convection reaches high in the atmosphere, so that the observation of atmospheric lithium through the Li-I line reflects the core abundance, since the evolutionary timescale is orders of magnitude larger than the convection timescale. Therefore, the observation of lithium in an
object older than $\sim$10$^8$ yrs - the timescale for the destruction of lithium
in the lowest mass stars -
signifies the lack of hydrogen burning.
A complementary outcome of the lithium test is the determination of the age of open clusters by finding the lithium-depletion boundary which separates objects without lithium from slightly fainter objects with lithium
(see e.g. Basri 1998, Stauffer et al. 1998).

As mentioned above, the timescale for the destruction of lithium in the lowest mass stars is $\sim$10$^8$ yr so that the lithium-test as a direct identification of the substellar nature of an object can not be applied to clusters younger than this age (although it can still be used to locate the substellar boundary in age or luminosity). Deuterium-burning, which occurs at a cooler temperature (see below) provides a new test at younger ages ($\simle 5\times 10^7$ yr, see \S2) to determine the age and/or mass of an object (B\'ejar, Zapatero Osorio \& Rebolo 1999).
From a more fundamental viewpoint, deuterium-burning has been suggested to play a key role in the formation of star-like objects (Shu, Adams \& Lizano 1987), following the collapse of a protostellar cloud, so that D-burning may define the minimum mass of brown dwarfs and thus of star-like object formation.

The aim of the present letter is to calculate the depletion of deuterium along evolution for different SSO masses, i.e. to derive [$^2$D]/[$^2$D$_0$]-age-mass relations, where [$^2$D$_0$]=$2\times 10^{-5}$ is the initial deuterium mass-fraction, and to relate this quantity to observable quantities, magnitude and color, in various optical and infrared passbands.
We also characterize the spectroscopic signature of deuterium in the spectra of these objects.

\section {Results}

The basic theory entering the present calculations has been described in detail elsewhere (Chabrier \& Baraffe 1997; Baraffe et al 1997, 1998). The atmosphere models are the so-called NextGen models (Hauschildt, Allard \& Baron 1999). Evolutionary models based on these inputs successfully reproduce observations in various diagrams (color-magnitude, color-color, mass-magnitude) for main-sequence stars (see references above) and pre-main-sequence stars (PMS) (Baraffe et al., 2000). The surface gravity $g$ of all the objects of interest in the present study (SSOs at $t\simle 10^8$ yr) is always larger than $g=10^3$ cm s$^{-2}$ so that the atmospheric structure is
not affected by sphericity effects. The central density and temperature are in the range $\rho_c\approx 1-100$ g cm$^{-3}$ and $T_C\approx 5\times 10^5-10^6$ K. Under these conditions correlation effects between ions in the plasma largely dominate the kinetic contribution and lower the Coulomb barrier between two fusing particles, yielding an enhancement of the reaction rate, as identified initially by Schatzman (1948) and Salpeter (1954). Moreover, for conditions characteristic of SSO interiors, the electron screening length is of the order of the interionic spacing, so that the ion-ion interaction is screened by the polarized electron background (Chabrier \& Baraffe 1997). For the mass-range
of interest, the electronic and ionic screening factors are indeed of 
comparable importance. The present calculations include such screening
corrections arising from both ion and electron polarization to thermonuclear reaction rates of D-burning through the $^2D(p,\gamma)^3He$ reaction.
The present calculations, based on the afore-outlined theory, represent significant improvements upon previous calculations of D-burning in low-mass stars, based in particular on approximate treatments of the atmosphere (e.g. D'Antona \& Mazzitelli, 1994). 

Calculations for PMS stars have been performed over the mass range $0.010 \,\msol \le M/\msol \le 1.2\,\msol$,
for solar metallicity, for ages $t\ge 10^6$ yr (Baraffe et al., 2000). In the present paper, we focus on objects near the D-burning limit, i.e. $0.012\,\msol \le M /\msol \le 0.1\,\msol$. At such young ages, central D-burning is followed by a rapid contraction of the object and thus a significant variation of the surface gravity within a few Myr (see Table 1).
For the characteristic range of effective temperatures $1800\,{\rm K}\simle \te \simle 3000$ K, convection reaches, or extends close to, the atmospheric layers where deuterated water forms. Therefore, the observed D-abundance should reflect the bulk abundance of the object. As shown recently by Chabrier et al. (2000) the core of massive and old brown dwarfs becomes dense and cold enough for electronic conduction to become the main energy transport mechanism. The onset of
conduction, however, occurs only for BDs older than 1 Gyr, and thus does not occur under conditions characteristic of the deuterium-test. 

As for hydrogen, the ignition temperature
translates into a deuterium-burning minimum mass (DBMM). With the afore-mentioned screening factors, these are found to be $T_D\simeq 4.0\times 10^5$ K and $m_D\simeq 0.012\,\msol$, in agreement with previous calculations (Saumon et al. 1996; Burrows et al. 1997).
Objects above the DBMM destroy deuterium on a very short time scale in the age range $\sim$ 4 - 50 Myr, for $m\sim 0.07-0.015\,\msol$, respectively,
whereas objects below this mass preserve their original deuterium abundance.

Table 1 displays the age for which
the initial D-abundance has been decreased by a factor 2 (50\% depletion) and 
100 (99\% depletion)
in the mass-range of interest, with the corresponding gravity, effective temperature, luminosity and absolute magnitudes. Figure 1 illustrates the D-depletion in the K-band with the corresponding effective temperature.
Superposed are the evolutionary tracks of several masses for SSOs, above the DBMM. As illustrated on this figure and Table 1, knowledge of deuterium-depletion provides a powerful determination of the mass of an object, for a given age. Conversely, the position of the 50\%-depletion boundary
(undistinguishable from the 0\%-depletion boundary), and the inferred magnitude, provide an independent method of age
determination for clusters younger than $\sim 50$ Myr, the deuterium-burning timescale of a 0.015 $\msol$ BD, just above the DBMM. It is worth noting that for an age $t\sim 1$ Myr, this boundary corresponds to the hydrogen-burning minimum mass, $m_{HBMM}\sim 0.07\,\msol$ (Chabrier \& Baraffe 1997), so that detection of deuterium in an object older than this age ensures its substellar nature. This provides a very useful diagnostic in
very young clusters where the lithium-test can not be applied.

Figure 2 displays the borderline for 50\% and 99\% D-depletion in a $L$-$\te$ diagram, with isochrones of 1 to 50 Myr. The crossing point between a depletion curve and an isochrone indicates the location of the (L,$\te$) point which corresponds to the corresponding amount of D-depletion for this age. If we choose the 99\%-depletion curve, objects cooler and fainter than the crossing point for a given age will all preserve deuterium.

The observable spectroscopic signature of D-depletion is a challenging task.
The most distinct features of the presence of deuterium occur in the range $\lambda = 2.8$ - 10 $\mu$m. Figure 3 shows the relative differences between a spectrum with deuterium set to cosmic abundance and a spectrum with full D-depletion in this wavelength-range for $\te=2500$ K, $\log g=3.5$ for a solar composition. As shown in the figure, spectral signatures of deuterium are in principle observable. Given the lack of accurate linelists, however, it is presently difficult to quantify exactly the amount of flux absorption due to the deuterium molecules transitions (HDO, OD). The present calculations were done using the most recent AMES linelists (Partridge \& Schwenke 1997). The situation should improve significantly in the near future and will yield more precise predictions. Spectral resolution of 30000 and very high signal-to-noise ratio are required to observe this signature (assuming rotation of the object is small enough). Such conditions are within reach with NIRSPEC on KeckII and with instruments planed for the VLT.
The observation of bands of deuterated methane CH$_3$D at 1.6 $\mu$m could also be used in principle to identify the presence of deuterium.
Methane, however, can not be used
for the deuterium test as an age-indicator:
at the temperature methane forms ($T\simle 1800$ K), SSOs above the D-burning minimum mass are old enough for all
 deuterium to have been burned.

\section{Conclusion}

In this {\it Letter}, we have derived deuterium-depletion abundances as a function of age for various masses in the substellar domain, and we have characterized the effective temperature, luminosity and observable quantities - absolute magnitudes and colors - corresponding to these depletion factors. We have also characterized the deuterium-burning minimum mass, temperature and luminosity as well as the spectral signatures of the presence of deuterium. These results are based on state-of-the-art, consistent  evolutionary calculations
between the interior structure, the atmosphere profile and the emergent spectrum. They provide the necessary ingredients to apply the deuterium test to identify the age or mass of young substellar objects and to provide an independent age determination of very young clusters. They also serve to identify objects at the deuterium-burning limit, as the lithium-test has been used to characterize objects at the hydrogen-burning limit. The unambiguous identification of free floating objects below this deuterium-burning minimum mass would indeed suggest that the formation of star-like objects extends below the deuterium-burning minimum mass.
Conclusions when applying this test, however, must be drawn with caution, for
exoplanets formed in multiple systems may be ejected from the system and end up
as free floating bodies.

\medskip

{\it Note}: For the transformation of observables into an $L$-$\te$ diagram, a consistent solution is to use the NextGen grid of synthetic spectra.
Spectral type vs $\te$ relations for young objects are not reliable and using such methods will yield results of dubious validity.
The NextGen grid is available by anonymous ftp. Please contact the authors for access details.

\medskip

{\it Acknowledgments}:
One of us (PH) acknowledges an invited visiting professor position
supported by the P\^ole
Scientifique de Mod\'elisation Num\'erique at ENS-Lyon. Part of this
work was supported by NSF grant
AST-9720704, NASA ATP grant NAG 5-8425 and LTSA grant NAG 5-3619, as
well as NASA/JPL grant 961582 to the University of Georgia, NASA LTSA
grant NAG5-3435 and NASA EPSCoR grant NCCS-168 to Wichita State
University. Some of the
calculations presented in this paper were performed on the CNUSC IBM
SP2, the IBM SP2 and SGI Origin 2000 of the UGA UCNS, on the IBM SP of
the San Diego Supercomputer Center (SDSC), with support from the
National Science Foundation, and on the Cray T3E of the NERSC with
support from the DoE.  We thank all these institutions for a generous
allocation of computer time.
\vfill
\eject

\section*{References}

\bib Baraffe, I., Chabrier, G., Allard, F., Hauschildt, P.H., 1997, \aap, 327, 1054
\bib Baraffe, I., Chabrier, G., Allard, F., Hauschildt, P.H., 1998, \aap, 337, 403
\bib Baraffe, I., Chabrier, G., Allard, F., Hauschildt, P.H., 2000, in preparation
\bib Basri, G., 1998, in {\it Cool stars in clusters and associations: magnetic activity and age indicators}, Mem. Soc. Astr. Ital., 68, 917
\bib Basri, G., Marcy, G.W.M. and Graham, J.R., 1996, \apj, 458, 600
\bib Basri, G., 2000, \araa, 38
\bib B\'ejar, V.J.S., Zapatero Osorio, M.R., Rebolo, R., 1999, \apj, 521, 671
\bib Burrows, A., Marley, M., Hubbard, W.B., Lunine, J.I., Guillot, T., et al. 1997, \apj, 491, 856
\bib Chabrier, G., \& Baraffe, I., 2000, \araa, 38, 339 
\bib Chabrier, G., \& Baraffe, I., 1997, \aap, 327, 1039
\bib Chabrier, G., Baraffe, I., Allard, F., Hauschildt, P.H., 2000, \apj, in press
\bib D'Antona, F., \& Mazzitelli, I., 1994, \apjs, 90, 467
\bib Hauschildt, P.H., Allard, F., Baron, E., 1999, \apj, 512, 377
\bib Partridge, H., \& Schwenke, D.W., 1997, {\it J. Chem. Phys.}, 106, 4618
\bib Rebolo, R., Mart\'\i n, E.L., Magazz\`u, A, 1992, \apjl, 389, 83
\bib Salpeter, E.E., 1954, {\it Austr. J. Phys.}, 7, 373
\bib Saumon, D., Hubbard, W.B., Burrows, A., Guillot, T., Lunine, J.I., Chabrier G., 1996,\apj, 460, 993
\bib Schatzman, E., 1948, {\it J. Phys. Rad.} 9, 46
\bib Shu, F.H., Adams, F.C., Lizano, S., 1987, \araa, 25, 23
\bib Stauffer, J.R., Schultz, G.,  Kirkpatrick, J.D., 1998, \apj, 499, L199

\vfill
\eject

\begin{table*}
\caption{Characteristic of the models when the initial D-abundance ($[D]_0=2\times 10^{-5}$) has been
depleted by a factor of 2 (upper row) and 100 (lower row),
for [M/H]=0.}				       
\begin{tabular}{cccccccccccc}
\hline\noalign{\smallskip} 
$m/\msol$ & $t$ (Myr) & $\te$ & $\log L/\lsun$ & $\log g$ & $M_B$ &$M_V$ &$M_R$ &$M_I$ &$M_J$ & $M_H$ & $M_K$ \\
\noalign{\smallskip}
\hline\noalign{\smallskip}
 0.015 & 18.42& 2177. & -3.17 &  4.09 & 20.62 & 19.14 & 17.17 & 14.47 & 10.41 &
 9.76 &  9.30 \\
&50.58& 1834. & -3.65 &  4.27 & 24.30 & 22.26 & 19.39 & 16.41 & 11.65 &
10.86 & 10.44 \\
 0.020 &  7.59& 2497. & -2.67 &  3.94 & 17.33 & 16.15 & 14.69 & 12.42 &  9.22 &
 8.61 &  8.19 \\
&17.02& 2415. & -2.88 &  4.10 & 18.39 & 17.08 & 15.52 & 13.13 &  9.72 &
 9.10 &  8.68\\
 0.030 &  2.92& 2703. & -2.21 &  3.79 & 15.22 & 14.07 & 12.74 & 10.77 &  8.12 &
 7.53 &  7.14 \\
&7.28& 2682. & -2.36 &  3.94 & 15.74 & 14.55 & 13.20 & 11.19 &  8.50 &
 7.91 &  7.52\\
 0.040 &  2.02& 2794. & -1.93 &  3.70 & 14.20 & 13.07 & 11.78 &  9.92 &  7.47 &
 6.87 &  6.50 \\
&4.99& 2784. & -2.08 &  3.84 & 14.62 & 13.46 & 12.17 & 10.29 &  7.82 &
 7.22 &  6.85\\
 0.050 &  1.55& 2844. & -1.74 &  3.64 & 13.55 & 12.44 & 11.18 &  9.37 &  7.00 &
 6.39 &  6.03 \\
&3.93& 2846. & -1.87 &  3.77 & 13.90 & 12.76 & 11.50 &  9.69 &  7.33 &
 6.73 &  6.37\\
 0.060 &  1.31& 2887. & -1.58 &  3.59 & 13.04 & 11.93 & 10.69 &  8.93 &  6.62 &
 6.01 &  5.65 \\
&3.27& 2887. & -1.72 &  3.72 & 13.38 & 12.25 & 11.02 &  9.24 &  6.95 &
 6.35 &  5.99\\
 0.070 &  1.13& 2920. & -1.45 &  3.54 & 12.62 & 11.53 & 10.30 &  8.56 &  6.30 &
 5.68 &  5.33 \\
&2.82& 2925. & -1.58 &  3.68 & 12.94 & 11.82 & 10.60 &  8.86 &  6.63 &
 6.02 &  5.67\\
 0.072 &  1.07& 2926. & -1.43 &  3.53 & 12.55 & 11.46 & 10.23 &  8.50 &  6.24 &
 5.62 &  5.27 \\
&2.74& 2932. & -1.56 &  3.67 & 12.86 & 11.74 & 10.53 &  8.79 &  6.57 &
 5.96 &  5.61\\
 0.075 &  1.03& 2932. & -1.39 &  3.52 & 12.45 & 11.36 & 10.14 &  8.41 &  6.15 &
 5.54 &  5.19 \\
&2.64& 2942. & -1.52 &  3.66 & 12.75 & 11.63 & 10.42 &  8.70 &  6.48 &
 5.87 &  5.53\\
  \hline
  \end{tabular}
  \end{table*}
\vfill
\eject
\onecolumn

\centerline {\bf FIGURE CAPTIONS}
\vskip1cm

\indent{\bf Figure 1 :}  Deuterium-depletion curves as a function of age
in the K-band.
The dashed lines correspond to different SSO masses, while
the solid lines correspond to the 50\% and 99\% D-depletion
limit, respectively. The inset displays the corresponding $\te(t)$.
\vskip1cm

\indent{\bf Figure 2 :}  $L$-$\te$ diagram for objects above the D-burning mass. Solid lines correspond to the 50\% (right curve) and 99\% (left curve) D-depletion
limit, respectively, dotted lines to isochrones (in Myr) and dashed lines to cooling sequences for several masses.
\vskip1cm

\indent{\bf Figure 3 :} Relative difference in per-cent between infrared spectra calculated
for $T_{\rm eff}=2500\,$K and $\log(g)=3.5$ models with 
complete deuterium depletion and no deuterium depletion $= (F_0-F_D)/(F_0+F_D)$,
where $F_D$ is the flux with solar D-abundance and $F_0$ is the flux with zero D-abundance.
The spectral resolution is about 30,000
(reduced from a computed resolution of about 300,000).
The vast majority of differences are due to HDO lines
seen on top of a background of H$_2$O lines. This
figure was generated using the AMES linelists
for water and HDO, which have the currently largest 
density of lines.
  
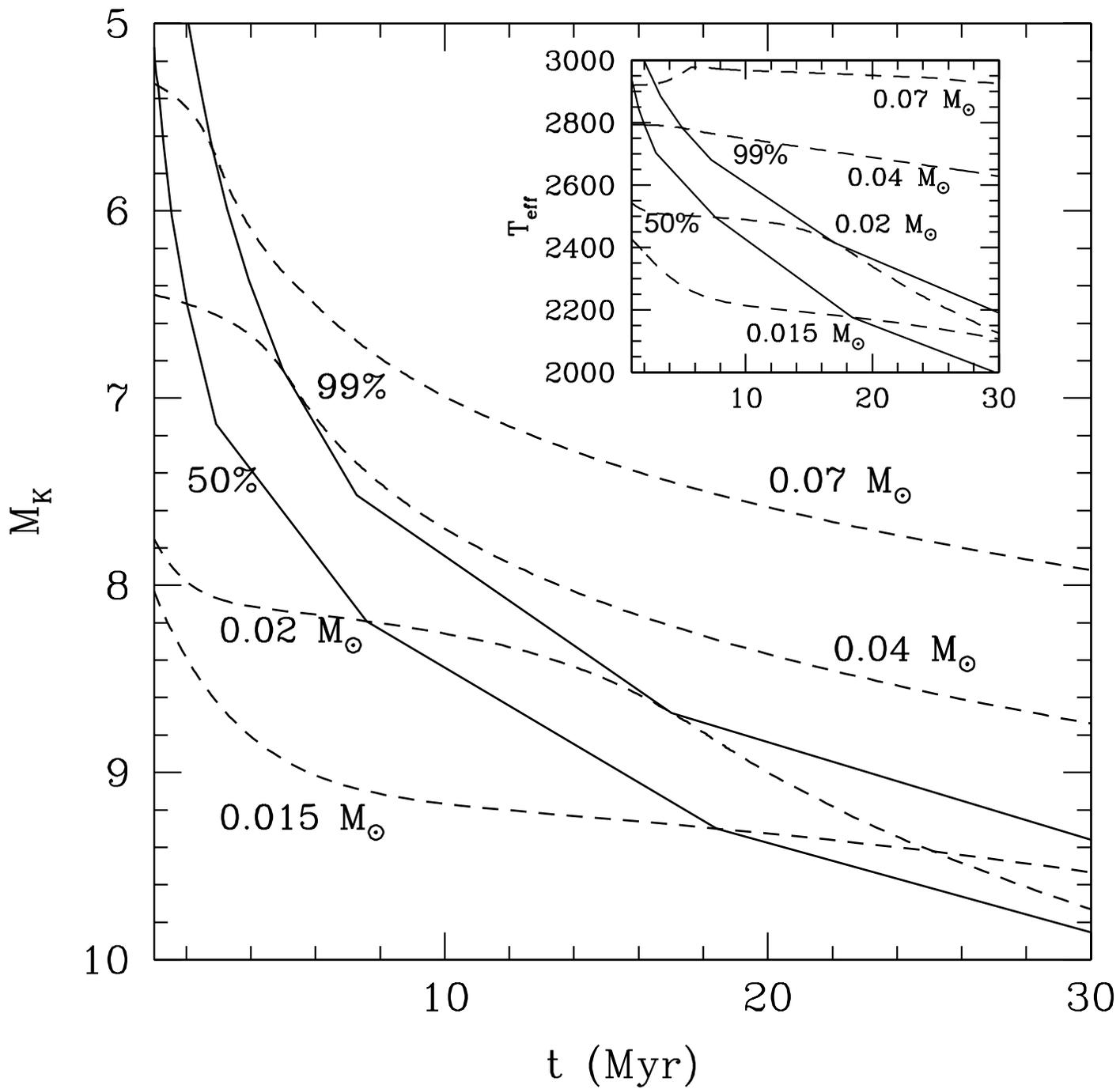
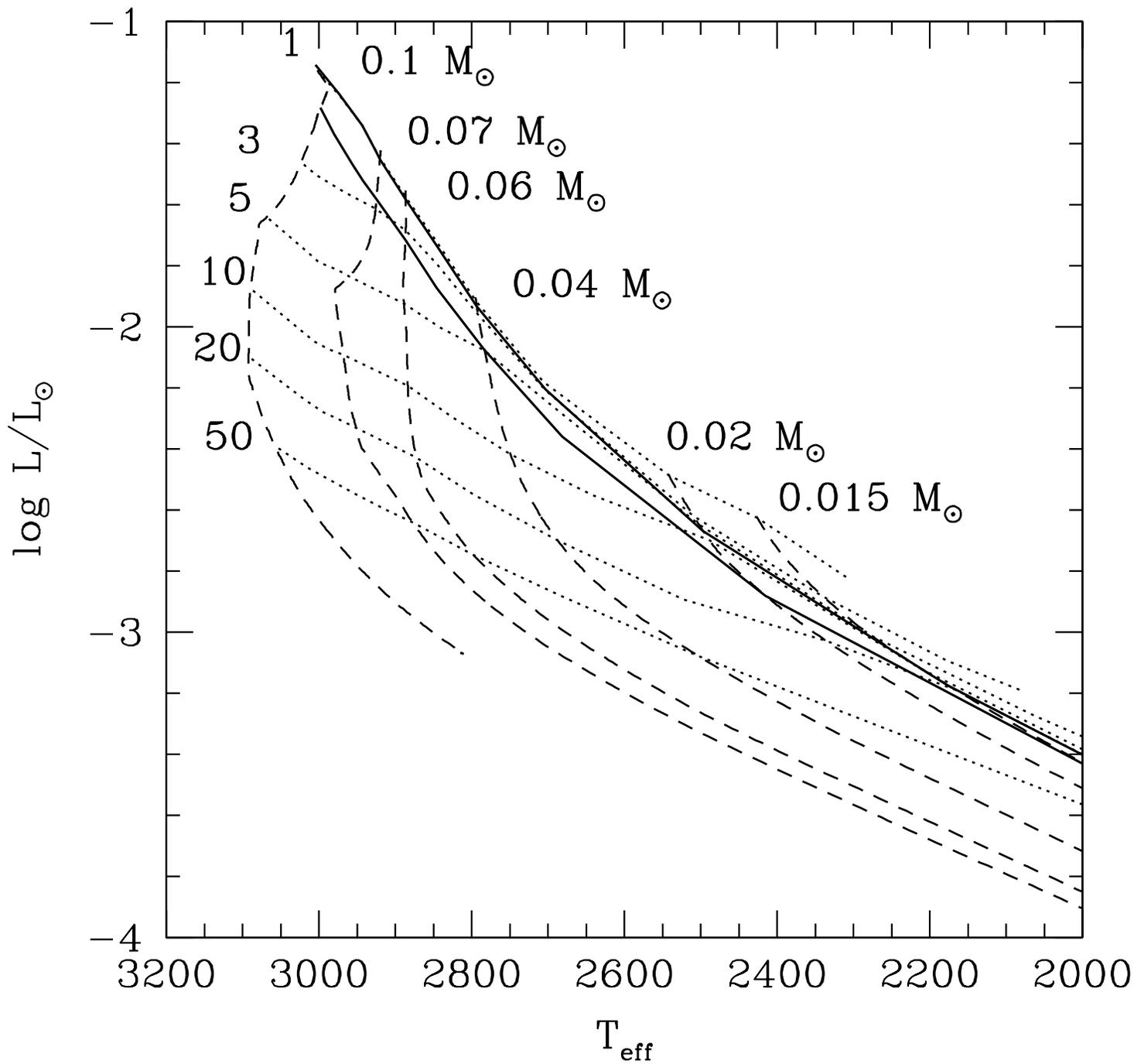
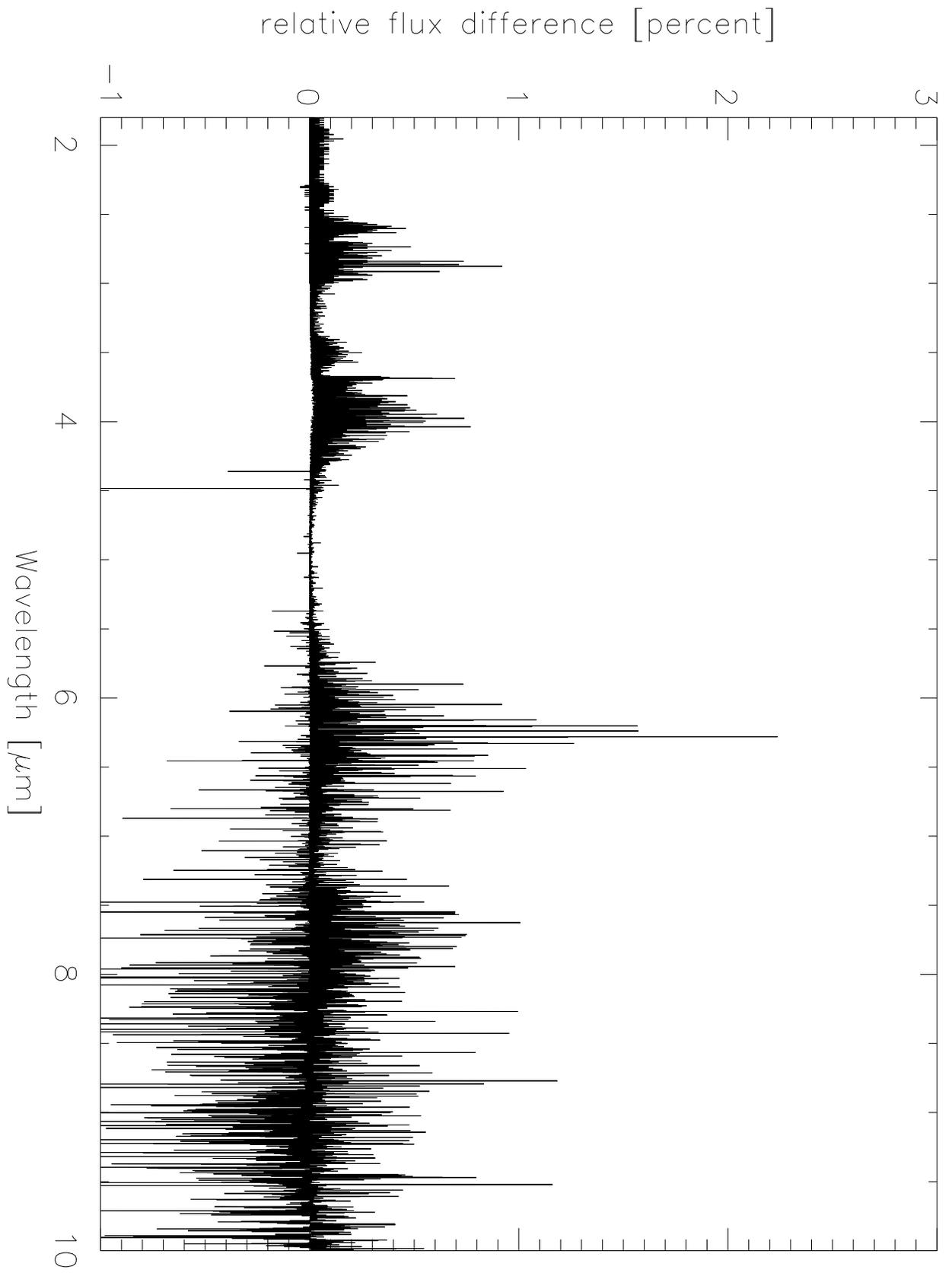
\end{document}